\documentclass[journal=jacsat,manuscript=article]{achemso}

\usepackage{chemformula} 
\usepackage[T1]{fontenc} 

\newcommand{\tablenotea}[1]{\parbox{15.5cm}{\indent \footnotesize{#1}}}

\author{Marcelino Ag\'undez}
\affiliation{Instituto de F\'isica Fundamental, CSIC, Calle Serrano 123, 28006 Madrid, Spain}
\email{marcelino.agundez@csic.es}
\author{Jos\'e Cernicharo}
\affiliation{Instituto de F\'isica Fundamental, CSIC, Calle Serrano 123, 28006 Madrid, Spain}
\email{jose.cernicharo@csic.es}

\title[TMC-1]{TMC-1: probing the onset of chemical complexity in space}

\abbreviations{IR,NMR,UV}


\begin{document}

\begin{abstract}

In recent years, the obsessive interest in the observation of TMC-1 has brought a boost in our knowledge of the chemistry of cold dark clouds. The number of molecules detected in this particular cloud has been more than doubled. Two observational programmes, GOTHAM and QUIJOTE, are responsible for this spectacular achievement. Here we provide an overall view of QUIJOTE, which is a line survey carried out in the Q band (31-50 GHz) with the Yebes\,40m radiotelescope, summarize the actual observational status of TMC-1, and discuss the chemistry of this remarkable source. We highlight the successes and failures of state-of-the-art chemical models to describe its chemical composition, with a particular emphasis on the origin of PAHs, which is yet far from being understood.

\end{abstract}

\begin{figure}[ht!]
\centering
\includegraphics[angle=0,width=1.0\textwidth]{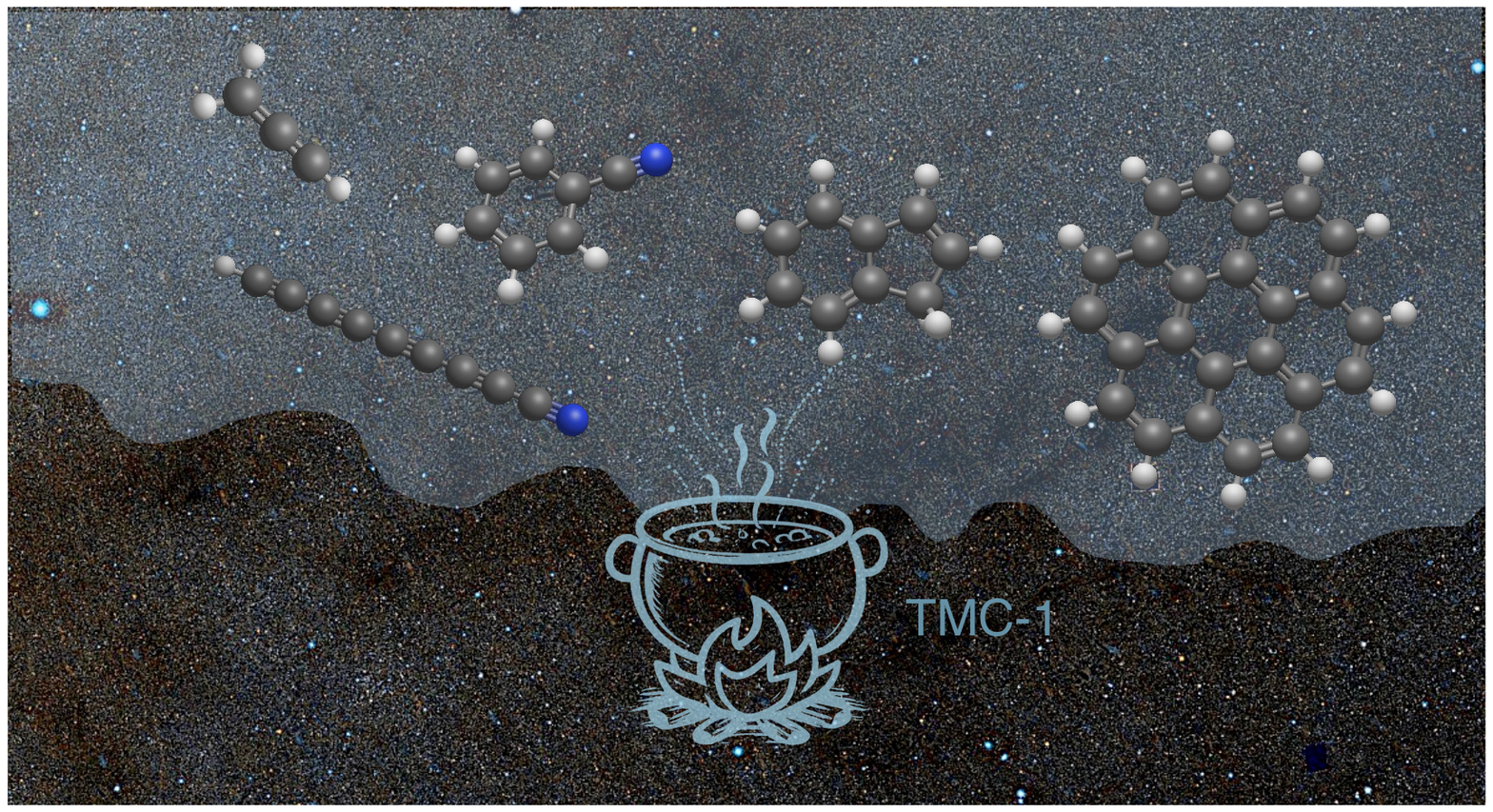} \label{TOC_graphic}
\end{figure}


\section{Introduction}

Cold dark clouds can be viewed as the places where the dawn of chemical complexity takes place. Indeed, the matter that forms these clouds is inherited from a previous stage of diffuse cloud \cite{Bergin2007}, where little chemical complexity is developed. Although the first transition from atoms to molecules occurs earlier, in the ejecta of evolved stars, most molecules formed in such winds are destroyed when matter becomes exposed to the interstellar ultraviolet (UV) field and do not reach further evolutionary stages \cite{Glassgold1996}. A notable exception concerns dust grains and large carbonaceous molecules such as polycyclic aromatic hydrocarbons (PAHs) and fullerenes, which are stable against UV radiation \cite{Kwok2009}. Therefore, cold dark clouds, where external UV radiation cannot penetrate, offer the first protected environment where chemical complexity can be safely developed. Such chemical complexity will be processed in further evolutionary stages related to protostar and protoplanetary disk formation, and will eventually end up in planetary systems, as evidenced from the observation of solar system bodies \cite{Altwegg2019}. Therefore, cold dark clouds are one of the best interstellar laboratories in which to study the first steps toward chemical complexity in space.

The Taurus Molecular Cloud 1 (commonly known as \mbox{TMC-1}) is a remarkable source in the sky for those researchers interested in astrochemistry, such as Prof. Eric Herbst. TMC-1 belongs to the Taurus Molecular Cloud complex, which is one of the best regions to study star formation. It is nearby, at a distance of 140 pc \cite{Elias1978}, and contains hundreds of dense molecular cores and newly formed stars \cite{Qian2012,Kenyon1995,Luhman2010}. TMC-1 appears as a ridge-shaped dark cloud inside Heiles Cloud 2 (HCL2) \citep{Heiles1968} extending 5'\,$\times$\,15' (0.2\,$\times$\,0.6 pc) along the southeast to northwest direction, with a position in the south-eastern side bright in emission of cyanopolyynes (the so-called cyanopolyyne peak) and a position in the north-western region, close to an IRAS infrared source thought to be a protostar, where \ch{NH3} emission is bright (the so-called ammonia peak). The cyanopolyyne peak is thought to be less evolved than the ammonia peak. The physical conditions at the cyanopolyyne peak of TMC-1 are those of a cold, dark, and dense molecular gas, with a temperature of 9 K, a visual extinction above 10 mag, and a volume density of 10$^4$ particles per cubic centimeter \citep{Agundez2023a,Fuente2019}, although its most remarkable characteristic is the exceptionally large variety of molecules observed. The chemical mechanism responsible for the presence of molecules in TMC-1 and other similar clouds was correctly identified by Prof. Eric Herbst in the early days of astrochemistry \cite{Herbst1973}. TMC-1 has been and continues to be profusely observed by radioastronomers and it has been also used as a template to test the validity of chemical models of cold dark clouds, where Prof. Eric Herbst has been the common thread through the years \cite{Leung1984,Herbst1986a,Herbst1986b,Millar1988,Herbst1989,Herbst1994,Smith2004,Wakelam2006,Wakelam2008,Walsh2009,Agundez2013}.

Here we revisit the state-of-the-art knowledge of the chemistry of TMC-1 in the light of the recent large observational campaigns, QUIJOTE and GOTHAM, that have more than doubled the number of molecules detected in this paradigmatic cloud. We discuss how well we understand the underlying chemical processes at work and which are the main open problems, which currently concern the formation of PAHs in these cold and dark environments.

\section{The QUIJOTE\footnote{\textbf{Q}-band \textbf{U}ltrasensitive \textbf{I}nspection \textbf{J}ourney to the \textbf{O}bscure \textbf{T}MC-1 \textbf{E}nvironment} line survey}

Sensitive line surveys are the best tool to unveil the molecular content of astronomical sources and to search for new molecules. A key element for carrying out a detailed analysis of line surveys is the availability of spectroscopic information of the already-known species, their isotopologues, and their vibrationally excited states (see, e.g., the detection in TMC-1 of C$_6$H in its $\nu_{11}$ vibrational state \cite{Cernicharo2023c}). The ability to identify the maximum possible number of lines coming from already known molecules is greatly aided in TMC-1 by the fact that the cloud is very cold and the lines are very narrow, minimising the possibility of line blending. This leaves the cleanest possible forest of unidentified lines, which opens up a chance to discover new molecules and get insights into the chemical composition. In these cases, the cloud under study becomes a real spectroscopic laboratory. With the help of devoted spectroscopic catalogs it is possible to characterize a molecule from a line survey. If the sensitivity of a line survey is large enough, then many unknown species will be discovered providing key information on the ongoing chemical processes in the cloud.

One of the main goals of the Nanocosmos\footnote{\texttt{https://nanocosmos.iff.csic.es/}} ERC synergy project has been the construction of high sensitivity receivers to be installed at the 40m radio telescope of the Yebes observatory (Spain). The motivation was the search for chemical complexity in interstellar and circumstellar clouds. Two main astronomical targets were selected for such purpose: the starless cold core TMC-1 \cite{Cernicharo2021b} and the carbon-rich Asymptotic Giant Branch (AGB) envelope IRC\,+10216 \cite{Pardo2022}. Although both objects have very different physical and chemical initial conditions, a significant numbers of long carbon chains (closed shell species, radicals, and anions) have been found in both sources with similar relative abundances. In addition, IRC\,+10216 is the archetypal C-rich AGB envelope, and there are reasons to expect that PAHs are formed in its wind \cite{Cherchneff1992}. Hence, a comparison of the molecular abundances in both objects could provide new insights on the chemical processes leading to the formation of PAHs in space. 

From an observational point of view, TMC-1 and IRC\,+10216 exhibit very different line profiles, requiring different observational and instrumental approaches. The whole instantaneous frequency coverage of the Yebes\,40m Q band provides a nominal spectral resolution of 38 kHz, but linewidths are very different in TMC-1 and IRC\,+10216. For TMC-1 we decided to maintain the nominal spectral resolution of 38 kHz, which provides a velocity resolution of 0.2-0.3 kms$^{-1}$ and a total of 5$\times$10$^5$ spectral channels in each polarization of the receiver. For IRC\,+10216, the data were smoothed to a frequency resolution of $\sim$0.2 MHz, which translates to a velocity resolution of 1.2-1.9 km s$^{-1}$, sufficient to resolve the $\sim$\,29 km s$^{-1}$ wide lines. The FFTS spectrometer used fitted the available funding for this instrumental development and the resulting spectral resolution met the scientific requirements of Nanocosmos.

\begin{figure}
\centering
\includegraphics[angle=0,width=0.95\textwidth]{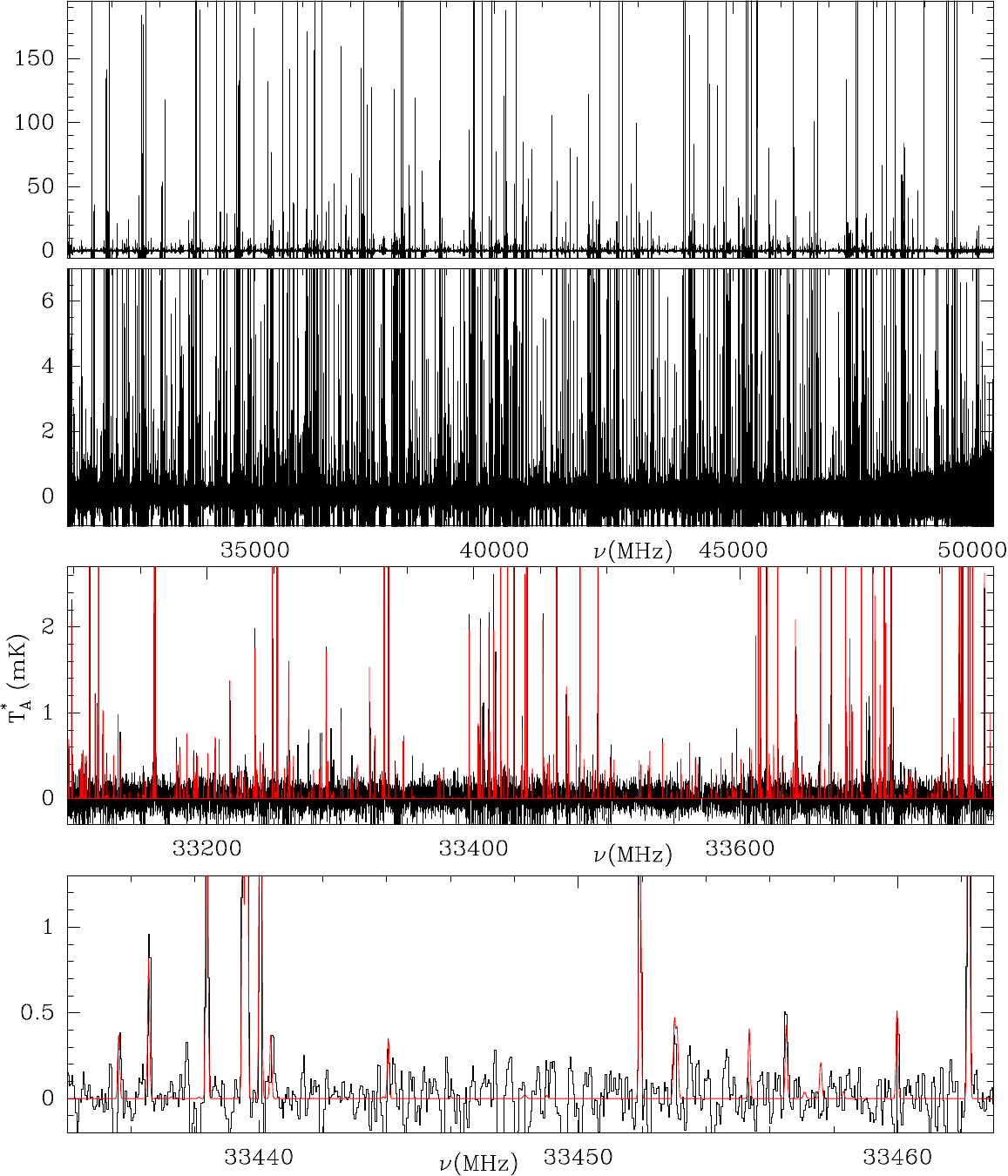}
\caption{{\small QUIJOTE line survey. The two upper panels show the whole survey at two different intensity scales. The two bottom panels show the survey with frequency ranges of 700 and 30 MHz. The red line in these bottom panels shows the synthetic spectrum computed using the molecular abundances given in Table\,\ref{table:molecules} and the physical parameters provided for each molecule in the corresponding reference quoted in the table. Negative features are produced in the folding of the frequency switching data of QUIJOTE. Any feature above 3\,$\sigma$ in the bottom panels without a red counterpart corresponds to an unidentified line.}} \label{fig:quijote}
\end{figure}

The scientific approach to chemical complexity in Nanocosmos is based on the standard method of line by line detection for each molecular species. In addition, our criteria for a reliable detection requires that all lines of the target molecule with expected emission above the 3\,$\sigma$ level in the survey must be detected. This means that we do not select just the lines that have a positive detected feature in the survey, but search for all the lines of the molecule lying in the covered frequency range. Any missing transition is analysed in detail and when there are more than 3 unexplained missing lines, the candidate molecule is considered as a non-detection. In view of the line density shown in Fig. \ref{fig:quijote}, detections based on one single line are not considered in QUIJOTE. This is a solid, robust and unambiguous procedure to detect molecules in space. It has been the main method for molecule detection since the beginning of astrochemistry in the 1970s. Alternative procedures, such as statistical analysis of the noise (line stacking and spectral matched filtering \cite{Loomis2018,Loomis2021}), have been used to provide detections by the GOTHAM project. Up to some level yet to be determined, most detections provided by this procedure, but no all \cite{Agundez2025b}, have been confirmed with QUIJOTE data. Moreover, some differences in the values derived for column densities and physical parameters, such as emission sizes (see below), emerge from both line surveys.

\begin{figure}[t]
\centering
\includegraphics[angle=0,width=0.85\textwidth]{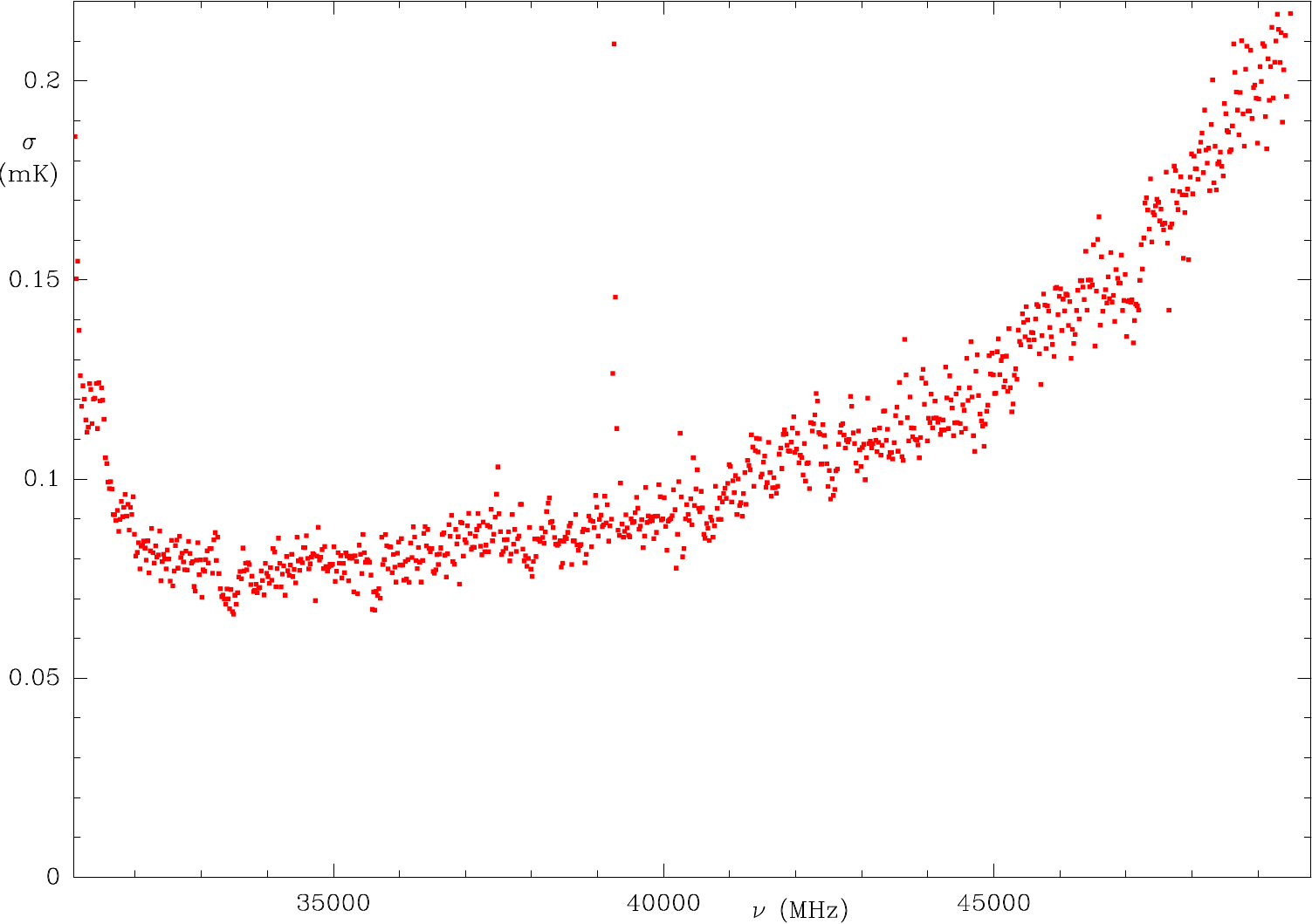}
\caption{{\small Measured sensitivity every 20 MHz of the QUIJOTE line survey (in milli Kelvin) as a function of the frequency. The spike around 39.2 GHz is due to radio frequency interferences and affects a frequency range of 80 MHz.}} \label{fig:sigma}
\end{figure}

The QUIJOTE line survey has been carried out between December 2019 and November 2024. The frequency range of the Q band, from 31.07 to 50.3 GHz, has been observed. Thanks to the design of the receivers and spectrometers of the new instrument, a total frequency coverage  across the Q band has been achieved in horizontal and vertical polarizations, which provides additional sensitivity for the final line survey. A detailed description of the telescope and receivers has been already provided \cite{Tercero2021}. A total of 1509 hours of observing time has been accumulated at the cyanopolyyne peak of TMC-1 ($\alpha_{J2000}=4^{\rm h} 41^{\rm  m} 41.9^{\rm s}$ and $\delta_{J2000}=+25^\circ 41' 27.0''$). The observing procedure was frequency switching with two different frequency throws of 8 and 10 MHz. The final sensitivity of QUIJOTE as a function of frequency is shown in Fig. \ref{fig:sigma}. It varies between $\sim$0.07 mK at 33 GHz and 0.2 mK at 50 GHz. The sensitivity of QUIJOTE surpasses any previous line survey \cite{Kaifu2004} by a factor 50-100. This sensitivity provides an unprecedented capacity to detect new molecules and of the isotopologues of the most abundant species. In this context, in the last 7 years QUIJOTE has permitted the discovery in space of around 70 new molecules \cite{Cernicharo2020a,Marcelino2020,Agundez2021b,Cabezas2021b,Cabezas2021c,Loru2023,Cernicharo2021a,Cernicharo2021b,Cernicharo2021c,Cernicharo2021d,Cernicharo2021e,Cernicharo2021f,Cernicharo2021g,Cernicharo2021h,Cernicharo2021i,Cernicharo2021j,Agundez2022b,Cabezas2022a,Cabezas2022b,Cabezas2022c,Cernicharo2022a,Cernicharo2022b,Cernicharo2022c,Fuentetaja2022a,Fuentetaja2022b,Agundez2023c,Cernicharo2023a,Fuentetaja2023,Marcelino2023,Silva2023,Cabezas2023,Cabezas2024a,Cabezas2024b,Fuentetaja2024a,Fuentetaja2024b,Agundez2024,Cernicharo2024a,Cernicharo2024b,Cernicharo2024c,Cernicharo2024d,Cernicharo2024e,Agundez2025a,Agundez2025b,Cabezas2025a,Cabezas2025b,Cabezas2025c,Cernicharo2026a,Esplugues2026,Fuentetaja2026}, while GOTHAM has discovered around 20 new species \cite{Xue2020,McGuire2020,McCarthy2021,Loomis2021,Lee2021a,Lee2021b,McGuire2021,Burkhardt2021a,Shingledecker2021,Siebert2022,Sita2022,Remijan2023,Wenzel2024,Wenzel2025a,Wenzel2025b,Remijan2025} (see next section and Table\,\ref{table:molecules}).

\begin{figure}[t]
\centering
\includegraphics[angle=0,width=0.95\textwidth]{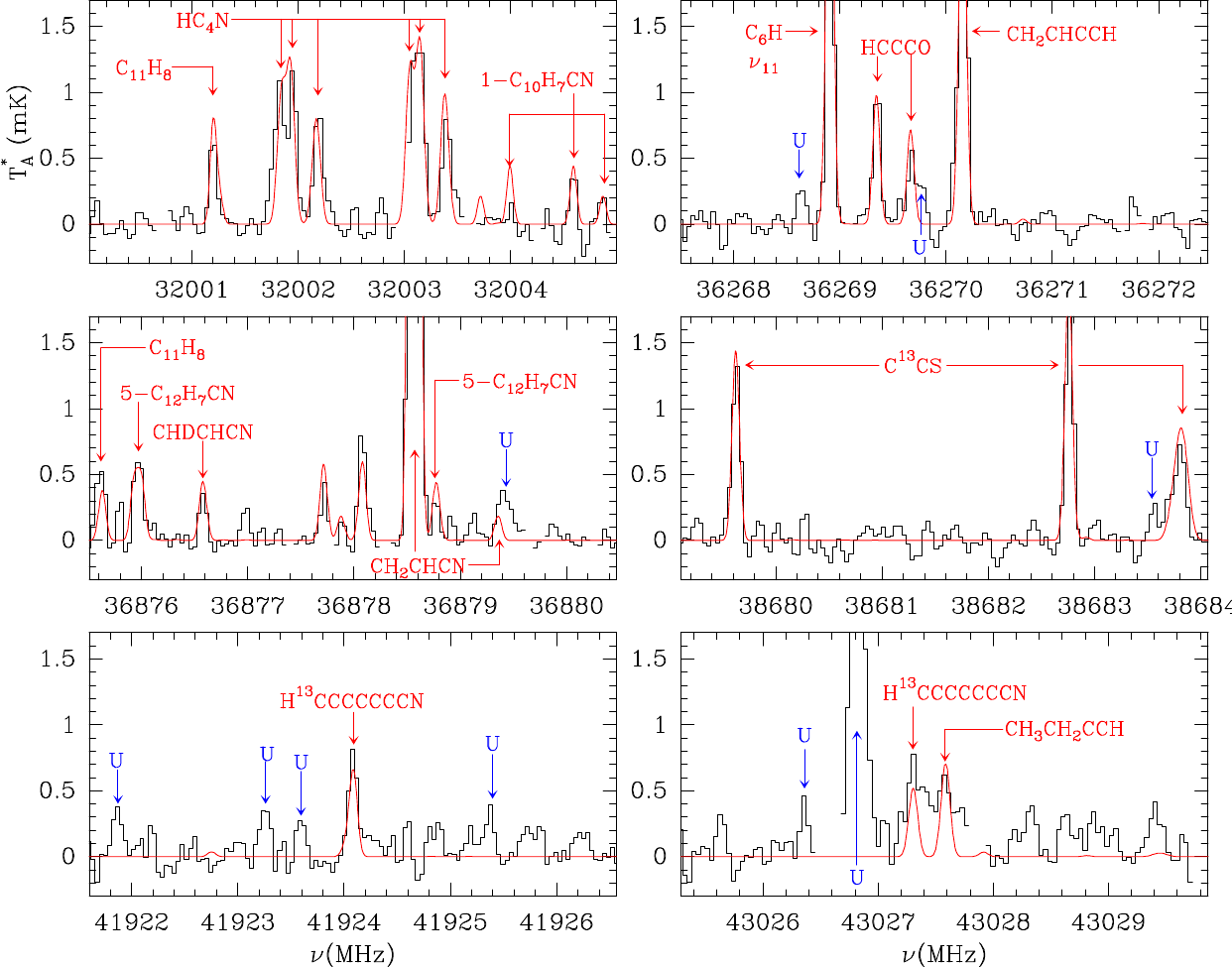}
\caption{{\small Selected spectral regions of QUIJOTE showing lines of key species detected through weak lines, such as hydrocarbons, carbon chain radicals, PAHs, vibrationally excited C$_6$H, and rare D and $^{13}$C isotopologues. These spectra show the unprecedented sensitivity and spectral quality of the QUIJOTE line survey. Unknown features above 3$\sigma$ are indicated in blue. Channels affected by the frequency switching
folding are blanked.}} \label{fig:exa_quijote}
\end{figure}

Figure\,\ref{fig:quijote} shows the line survey zoomed in on different intensity and frequency scales, showing that there is a spectacular forest of weak lines. Figure\,\ref{fig:exa_quijote} shows selected frequency windows over a frequency scale 5 MHz wide, where lines are identified. Initially most of these features were unknown, even when using our MADEX catalogue \cite{Cernicharo2012} which contained around 6000 spectral entries by 2020. Nowadays it contains 6742 species that cover all new molecules and isotopologues discovered with QUIJOTE, together with those detected prior to 2020 and  several thousand of potential species that have been incorporated in the code since 1985. Figure\,\ref{fig:exa_quijote} shows the presence of several PAHs such as \ch{C11H8} (1H-cyclopent[cd]indene) \cite{Fuentetaja2026}, \ch{C12H7CN} (cyanoacenaphthylene) \cite{Cernicharo2024a,Cernicharo2026a}, and \ch{C10H7CN} (cyanonaphthalene) \cite{McGuire2021,Cernicharo2024e}. In addition, exotic species detected previously only in C-rich stars such as HC$_4$N \cite{Cernicharo2004} are prominent in the QUIJOTE data (top left panel of Fig.\,\ref{fig:exa_quijote}).

The modeling of the observed emission in a line survey is often tackled with a very limited information of the spatial extent of the emission. In GOTHAM and QUIJOTE, the only available spatial information is provided by the variation of the telescope half power beam with the frequency across the line survey. While GOTHAM fits four velocity components with different spatial sizes, QUIJOTE assumes a single velocity component and a source radius of 40$''$ based on previous observations of TMC-1 in several molecular species \cite{Fosse2001,Lique2006}. None of these methods is satisfactory to obtain accurate column densities, which are absolutely needed to put constraints for chemical models of the source. 

\begin{figure}[t]
\centering
\includegraphics[width=0.95\textwidth,angle=0]{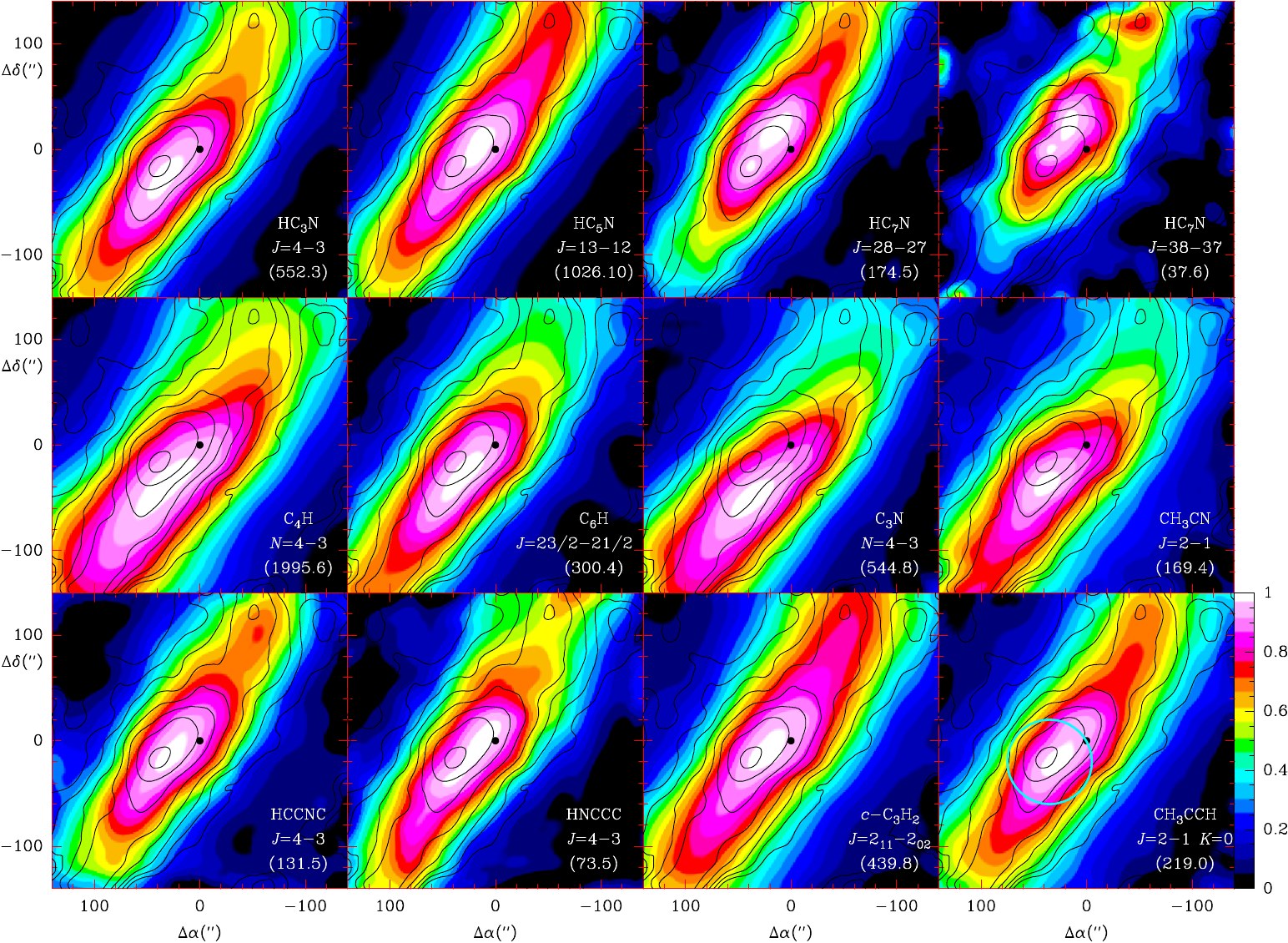}
\caption{{\small Integrated intensity between 5.3 and 6.5 kms$^{-1}$ of different molecular transitions (colors) compared with that of C$_6$H$_5$CN (black contours; first contour and step are 0.75 mK kms$^{-1}$). For each molecular transition the integrated intensity has been normalized to the maximum value within the area covered by the map. Hence, the color scale is the same for all molecular transitions and is indicated by the wedge at the lower right part of the Figure. The molecule, transition, and maximum intensity (shown between parentheses) are indicated at the lower right corner of each panel. The black dot corresponds to the centre of the map. The cyan circle of 40$''$ radius in the bottom right panel represents the emission size of molecules. The figure is from Ref.~\citenum{Cernicharo2023a}.}}
\label{fig:maps}
\end{figure}

The spatial distribution of the observed molecules, together with the issues related to the line opacities and radiative transfer, can only be addressed through spatial mapping of the molecular emission in different rotational transitions of each molecule. To overcome these issues, the QUIJOTE line survey has been complemented with high sensitivity maps obtained with the Yebes 40m radio telescope and covering a region of 320$''$\,$\times$\,320$''$ centered on the QUIJOTE position and covering the entire Q band. These maps are called SANCHO\footnote{\textbf{S}urveying the \textbf{A}rea of the \textbf{N}eighbour TMC-1 \textbf{C}loud through \textbf{H}eterodyne \textbf{O}bservations}, and are a faithful companion to the QUIJOTE line survey\cite{Cernicharo2023b}. They have a large sensitivity that has allowed the detection of species with low intensity lines such as benzonitrile and many other molecular species\cite{Cernicharo2023b}. Fig. \ref{fig:maps} shows the spatial distribution of \ch{C6H5CN} (black contours) superimposed on the spatial distribution (in colors) of the cyanopolyynes \ch{HC3N}, \ch{HC5N}, \ch{HC7N}, the radicals \ch{C4H}, \ch{C6H}, \ch{C3N}, and several selected carbon species \cite{Cernicharo2023b}. This Figure clearly shows the spatial coexistence of benzonitrile and other molecules along the TMC-1 filament. This strongly suggests a bottom-up mechanism for the formation of all these large carbon species, including large PAHs (see next sections).

Finally, the QUIJOTE line survey has been recently complemented with data gathered with a new receiver covering the frequency range 18 to 32 GHz (The K/Ka band) \cite{LopezPerez2026}. The first astronomical results concern the detection of two new cyano derivatives of acenaphthylene \cite{Cernicharo2026a}. The survey is still ongoing and we expect to have a similar sensitivity to that of QUIJOTE in the next 2-3 years.

\section{The chemistry of TMC-1} \label{sec:chemistry}

In Table\,\ref{table:molecules} we list all the molecules detected in TMC-1 with their associated column densities and the corresponding reference. Similar compilations based exclusively on Nobeyama \cite{Gratier2016} or GOTHAM \cite{Xue2025} line surveys are available. To date we count 188 molecules detected in TMC-1. This is a rather large number, taking into account that the total number of molecules discovered to date in interstellar or circumstellar media is around 340 \cite{Endres2016}. In fact, TMC-1 is probably the astronomical source with the richest diversity of molecules characterized so far. It is worth to note that around 60\,\% of the molecules known to be present in TMC-1 have been discovered in the last five years. This boom of detections has brought an unprecedented level of detail in our knowledge of the chemical composition of cold dark clouds. An obvious question that emerges is how well we understand the chemical processes behind the synthesis of this rich diversity of molecules observed.

\begin{table}
\scriptsize
\caption{Inventory of molecules detected in TMC-1 as of April 2026.}
\label{table:molecules}
\centering
\begin{tabular}{lcrlcrlcr}
\hline \hline
Molecule & $N$ (cm$^{-2}$) & Ref. & Molecule & $N$ (cm$^{-2}$) & Ref. & Molecule & $N$ (cm$^{-2}$) & Ref. \\
\hline
 CH                             & 2.0\,$\times$\,10$^{14}$ & \cite{Ohishi1992} &  HCOOCH$_3$                     & 1.1\,$\times$\,10$^{12}$ & \cite{Agundez2021a} &  HNC$_5$                        & 1.3\,$\times$\,10$^{10}$ & \cite{Fuentetaja2024b} \\
 C$_2$H                         & 6.5\,$\times$\,10$^{14}$ & \cite{Sakai2010} &  C$_3$O                         & 1.2\,$\times$\,10$^{12}$ & \cite{Cernicharo2021f} &  HC$_5$NH$^+$                   & 7.5\,$\times$\,10$^{11}$ & \cite{Marcelino2020} \\
 $l$-C$_3$H                     & 1.2\,$\times$\,10$^{13}$ & \cite{Cernicharo2026b} &  HC$_3$O                        & 1.3\,$\times$\,10$^{11}$ & \cite{Cernicharo2021f} &  2-C$_4$H$_5$CN                 & 3.1\,$\times$\,10$^{10}$\,$^a$ & \cite{Agundez2025b} \\
 $c$-C$_3$H                     & 1.3\,$\times$\,10$^{13}$ & \cite{Cernicharo2026b} &  HC$_3$O$^+$                    & 2.1\,$\times$\,10$^{11}$ & \cite{Cernicharo2020a} &  CHCCHCHCN                      & 3.0\,$\times$\,10$^{11}$ & \cite{Lee2021a} \\
 C$_3$H$^+$                     & 2.4\,$\times$\,10$^{10}$ & \cite{Cernicharo2022a} &  HCCCHO                         & 9.2\,$\times$\,10$^{11}$ & \cite{Esplugues2026} &  CH$_2$CHC$_3$N                 & 2.0\,$\times$\,10$^{11}$ & \cite{Lee2021a} \\
 $c$-C$_3$H$_2$                 & 1.2\,$\times$\,10$^{14}$ & \cite{Cabezas2021a} &  $c$-C$_3$H$_2$O                & 3.2\,$\times$\,10$^{11}$ & \cite{Esplugues2026} &  NC$_4$NH$^+$                   & 1.1\,$\times$\,10$^{10}$ & \cite{Agundez2023c} \\
 $l$-C$_3$H$_2$                 & 1.9\,$\times$\,10$^{12}$ & \cite{Cabezas2021b} &  $l$-H$_2$C$_3$O                & 3.7\,$\times$\,10$^{10}$ & \cite{Esplugues2026} &  $Z$-NCCHCHCN                   & 5.1\,$\times$\,10$^{10}$ & \cite{Agundez2024} \\
 CH$_2$CCH                      & 1.0\,$\times$\,10$^{14}$ & \cite{Agundez2022a} &  C$_2$H$_3$CHO                  & 2.2\,$\times$\,10$^{11}$ & \cite{Agundez2021a} &  CH$_3$C$_5$N                   & 9.5\,$\times$\,10$^{10}$ & \cite{Fuentetaja2022a} \\
 CH$_2$CCH$^+$                  & 7.0\,$\times$\,10$^{11}$ & \cite{Silva2023} &  CH$_3$CHCO                     & 1.5\,$\times$\,10$^{11}$ & \cite{Fuentetaja2023} &  CH$_2$CCHC$_3$N                & 1.2\,$\times$\,10$^{11}$ & \cite{Fuentetaja2022a} \\
 CH$_3$CCH                      & 1.5\,$\times$\,10$^{14}$ & \cite{Cernicharo2026b} &  CH$_3$COCH$_3$                 & 1.4\,$\times$\,10$^{11}$ & \cite{Agundez2023b} &  1-$c$-C$_5$H$_5$CN             & 3.1\,$\times$\,10$^{11}$ & \cite{Cernicharo2021e} \\
 CH$_2$CHCH$_3$                 & 4.0\,$\times$\,10$^{13}$ & \cite{Marcelino2007} &  C$_2$H$_5$CHO                  & 1.9\,$\times$\,10$^{11}$ & \cite{Agundez2023b} &  2-$c$-C$_5$H$_5$CN             & 1.3\,$\times$\,10$^{11}$ & \cite{Cernicharo2021e} \\
 C$_4$H                         & 8.5\,$\times$\,10$^{13}$ & \cite{Agundez2023a} &  C$_5$O                         & 1.5\,$\times$\,10$^{10}$ & \cite{Cernicharo2021f} &  C$_7$N$^-$                     & 5.0\,$\times$\,10$^{10}$ & \cite{Cernicharo2023a} \\
 C$_4$H$^-$                     & 2.1\,$\times$\,10$^{10}$ & \cite{Agundez2023a} &  HC$_5$O                        & 1.4\,$\times$\,10$^{12}$ & \cite{Cernicharo2021f} &  HC$_7$N                        & 2.1\,$\times$\,10$^{13}$ & \cite{Cabezas2022c} \\
 H$_2$C$_4$                     & 3.3\,$\times$\,10$^{12}$ & \cite{Cabezas2021b} &  HC$_7$O                        & 6.5\,$\times$\,10$^{11}$ & \cite{Cernicharo2021f} &  HC$_7$N$^+$                    & 2.3\,$\times$\,10$^{10}$ & \cite{Cernicharo2024d} \\
\cline{4-6}
 CH$_2$CHCCH                    & 9.5\,$\times$\,10$^{12}$ & \cite{Cernicharo2024a} &  NH$_3$                         & 5.0\,$\times$\,10$^{14}$ & \cite{Gratier2016} &  HC$_7$NH$^+$                   & 5.5\,$\times$\,10$^{10}$ & \cite{Cabezas2022c} \\
 CH$_3$CH$_2$CCH                & 6.2\,$\times$\,10$^{11}$ & \cite{Cernicharo2024a} &  N$_2$H$^+$                     & 2.8\,$\times$\,10$^{12}$ & \cite{Pratap1997} &  $c$-C$_6$H$_5$CN               & 1.2\,$\times$\,10$^{12}$ & \cite{Cernicharo2021e} \\
 C$_5$H                         & 1.3\,$\times$\,10$^{12}$ & \cite{Cabezas2022a} &  CN                             & 7.4\,$\times$\,10$^{12}$ & \cite{Pratap1997} &  CH$_3$C$_7$N                   & 8.6\,$\times$\,10$^{10}$ & \cite{Siebert2022} \\
 $c$-C$_5$H                     & 9.0\,$\times$\,10$^{10}$ & \cite{Cabezas2022a} &  HCN                            & 1.1\,$\times$\,10$^{14}$ & \cite{Pratap1997} &  HC$_9$N                        & 2.2\,$\times$\,10$^{13}$ & \cite{Loomis2021} \\
 C$_5$H$^+$                     & 8.8\,$\times$\,10$^{10}$ & \cite{Cernicharo2022a} &  HNC                            & 2.6\,$\times$\,10$^{14}$ & \cite{Pratap1997} &  2-$c$-C$_9$H$_7$CN             & 2.1\,$\times$\,10$^{11}$ & \cite{Sita2022} \\
 H$_2$C$_5$                     & 1.8\,$\times$\,10$^{10}$ & \cite{Cabezas2021b} &  HCNH$^+$                       & 1.9\,$\times$\,10$^{13}$ & \cite{Schilke1991} &  HC$_{11}$N                     & 7.8\,$\times$\,10$^{11}$ & \cite{Loomis2021} \\
 $c$-C$_3$HCCH                  & 3.1\,$\times$\,10$^{11}$ & \cite{Cernicharo2021d} &  NO                             & 2.7\,$\times$\,10$^{14}$ & \cite{Gerin1993} &  1-$c$-C$_{10}$H$_7$CN          & 7.4\,$\times$\,10$^{11}$ & \cite{McGuire2021} \\
 CH$_3$C$_4$H                   & 6.5\,$\times$\,10$^{12}$ & \cite{Cernicharo2021c} &  HCCN                           & 4.4\,$\times$\,10$^{11}$ & \cite{Cernicharo2021a} &  2-$c$-C$_{10}$H$_7$CN          & 7.1\,$\times$\,10$^{11}$ & \cite{McGuire2021} \\
 CH$_2$CCHCCH                   & 1.2\,$\times$\,10$^{13}$ & \cite{Cernicharo2021c} &  CH$_2$CN                       & 1.4\,$\times$\,10$^{13}$ & \cite{Cabezas2021a} &  1-$c$-C$_{12}$H$_7$CN          & 9.5\,$\times$\,10$^{11}$ & \cite{Cernicharo2024e} \\
 HCCCH$_2$CCH                   & 5.0\,$\times$\,10$^{12}$ & \cite{Fuentetaja2024a} &  CH$_3$CN                       & 4.7\,$\times$\,10$^{12}$ & \cite{Cabezas2021a} &  3-$c$-C$_{12}$H$_7$CN          & 7.0\,$\times$\,10$^{11}$ & \cite{Cernicharo2026a} \\
 $c$-C$_5$H$_6$                 & 1.2\,$\times$\,10$^{13}$ & \cite{Cernicharo2021d} &  CH$_3$NC                       & 3.0\,$\times$\,10$^{11}$ & \cite{Tennis2023} &  4-$c$-C$_{12}$H$_7$CN          & 5.0\,$\times$\,10$^{11}$ & \cite{Cernicharo2026a} \\
 C$_6$H                         & 4.8\,$\times$\,10$^{12}$ & \cite{Agundez2023a} &  NCO                            & 7.4\,$\times$\,10$^{11}$ & \cite{Cernicharo2024b} &  5-$c$-C$_{12}$H$_7$CN          & 9.5\,$\times$\,10$^{11}$ & \cite{Cernicharo2024e} \\
 C$_6$H$^-$                     & 1.5\,$\times$\,10$^{11}$ & \cite{Agundez2023a} &  HNCO                           & 1.1\,$\times$\,10$^{13}$ & \cite{Cernicharo2024b} &  1-$c$-C$_{16}$H$_9$CN          & 1.5\,$\times$\,10$^{12}$ & \cite{Wenzel2024} \\
 H$_2$C$_6$                     & 8.0\,$\times$\,10$^{10}$ & \cite{Cabezas2021b} &  HCNO                           & 7.8\,$\times$\,10$^{10}$ & \cite{Cernicharo2024b} &  2-$c$-C$_{16}$H$_9$CN          & 8.4\,$\times$\,10$^{11}$ & \cite{Wenzel2025a} \\
 HCCCHCCC                       & 1.3\,$\times$\,10$^{11}$ & \cite{Fuentetaja2022b} &  HOCN                           & 1.5\,$\times$\,10$^{11}$ & \cite{Cernicharo2024b} &  4-$c$-C$_{16}$H$_9$CN          & 1.3\,$\times$\,10$^{12}$ & \cite{Wenzel2025a} \\
 $o$-C$_6$H$_4$                 & 5.0\,$\times$\,10$^{11}$ & \cite{Cernicharo2021b} &  H$_2$NCO$^+$                   & 1.8\,$\times$\,10$^{10}$ & \cite{Cernicharo2024b} &  C$_{24}$H$_{11}$CN             & 2.7\,$\times$\,10$^{12}$ & \cite{Wenzel2025b} \\
\cline{7-9}
 C$_7$H                         & 6.5\,$\times$\,10$^{10}$ & \cite{Cernicharo2022b} &  CH$_2$CHCN                     & 6.2\,$\times$\,10$^{12}$ & \cite{Cernicharo2024a} &  H$_2$S                         & 2.2\,$\times$\,10$^{13}$ & \cite{Navarro-Almaida2020} \\
 CH$_3$C$_6$H                   & 7.0\,$\times$\,10$^{11}$ & \cite{Fuentetaja2022a} &  C$_3$N                         & 1.2\,$\times$\,10$^{13}$ & \cite{Agundez2023a} &  HS$_2$                         & 5.7\,$\times$\,10$^{11}$ & \cite{Esplugues2025} \\
 CH$_2$CCHC$_4$H                & 2.2\,$\times$\,10$^{12}$ & \cite{Fuentetaja2022a} &  C$_3$N$^-$                     & 6.4\,$\times$\,10$^{10}$ & \cite{Agundez2023a} &  CS                             & 1.1\,$\times$\,10$^{14}$ & \cite{Fuentetaja2025} \\
 $c$-C$_5$H$_4$CCH$_2$          & 2.7\,$\times$\,10$^{12}$ & \cite{Cernicharo2022b} &  HC$_3$N                        & 2.3\,$\times$\,10$^{14}$ & \cite{Cernicharo2020b} &  HCS$^+$                        & 5.5\,$\times$\,10$^{12}$ & \cite{Fuentetaja2025} \\
 1-$c$-C$_5$H$_5$CCH            & 1.4\,$\times$\,10$^{12}$ & \cite{Cernicharo2021e} &  HCCNC                          & 3.0\,$\times$\,10$^{12}$ & \cite{Cernicharo2020b} &  HCS                            & 5.5\,$\times$\,10$^{12}$ & \cite{Cernicharo2021i} \\
 2-$c$-C$_5$H$_5$CCH            & 2.0\,$\times$\,10$^{12}$ & \cite{Cernicharo2021e} &  HCCNCH$^+$                     & 3.0\,$\times$\,10$^{10}$ & \cite{Agundez2022b} &  HSC                            & 1.3\,$\times$\,10$^{11}$ & \cite{Cernicharo2021i} \\
 C$_8$H                         & 3.0\,$\times$\,10$^{11}$ & \cite{Agundez2023a} &  HNC$_3$                        & 5.2\,$\times$\,10$^{11}$ & \cite{Cernicharo2020b} &  H$_2$CS                        & 3.7\,$\times$\,10$^{13}$ & \cite{Fuentetaja2025} \\
 C$_8$H$^-$                     & 2.0\,$\times$\,10$^{10}$ & \cite{Agundez2023a} &  HC$_3$N$^+$                    & 6.0\,$\times$\,10$^{10}$ & \cite{Cabezas2024b} &  CH$_3$SH                       & 1.7\,$\times$\,10$^{12}$ & \cite{Agundez2025a} \\
 $c$-C$_6$H$_5$CCH              & 3.0\,$\times$\,10$^{12}$ & \cite{Loru2023} &  HC$_3$NH$^+$                   & 1.0\,$\times$\,10$^{12}$ & \cite{Marcelino2020} &  SO                             & 1.0\,$\times$\,10$^{14}$ & \cite{Lique2006} \\
 $c$-C$_9$H$_8$                 & 1.6\,$\times$\,10$^{13}$ & \cite{Cernicharo2021d} &  H$_2$CCCN                      & 2.5\,$\times$\,10$^{11}$ & \cite{Cabezas2023} &  HSO                            & 7.0\,$\times$\,10$^{10}$ & \cite{Marcelino2023} \\
 C$_{10}$H                      & 2.0\,$\times$\,10$^{11}$\,$^a$ & \cite{Remijan2023} &  CH$_3$CH$_2$CN                 & 1.3\,$\times$\,10$^{11}$ & \cite{Cernicharo2024a} &  NS                             & 1.7\,$\times$\,10$^{12}$ & \cite{Cernicharo2018} \\
 C$_{10}$H$^-$                  & 4.0\,$\times$\,10$^{11}$ & \cite{Remijan2023} &  CNCN                           & 8.0\,$\times$\,10$^{11}$ & \cite{Agundez2023c} &  NS$^+$                         & 5.2\,$\times$\,10$^{10}$ & \cite{Cernicharo2018} \\
 $c$-C$_{11}$H$_8$              & 6.0\,$\times$\,10$^{12}$ & \cite{Fuentetaja2026} &  NCCNH$^+$                      & 8.6\,$\times$\,10$^{10}$ & \cite{Agundez2015} &  C$_2$S                         & 3.4\,$\times$\,10$^{13}$ & \cite{Fuentetaja2025} \\
 $c$-C$_{13}$H$_{10}$           & 2.8\,$\times$\,10$^{13}$ & \cite{Cabezas2025c} &  HCOCN                          & 3.5\,$\times$\,10$^{11}$ & \cite{Cernicharo2021h} &  HC$_2$S$^+$                    & 1.1\,$\times$\,10$^{12}$ & \cite{Cabezas2022b} \\
\cline{1-3}
 OH                             & 3.0\,$\times$\,10$^{15}$ & \cite{Ohishi1992} &  HC$_4$N                        & 3.7\,$\times$\,10$^{11}$ & \cite{Cernicharo2021a} &  HC$_2$S                        & 6.8\,$\times$\,10$^{11}$ & \cite{Cernicharo2021i} \\
 CO                             & 1.7\,$\times$\,10$^{18}$ & \cite{Pratap1997} &  CH$_2$C$_3$N                   & 1.6\,$\times$\,10$^{11}$ & \cite{Cabezas2021c} &  H$_2$C$_2$S                    & 7.8\,$\times$\,10$^{11}$ & \cite{Cernicharo2021i} \\
 HCO                            & 1.1\,$\times$\,10$^{12}$ & \cite{Cernicharo2021f} &  HC$_3$HCN                      & 2.2\,$\times$\,10$^{11}$ & \cite{Cabezas2025a} &  OCS                            & 2.2\,$\times$\,10$^{13}$ & \cite{Matthews1987} \\
 HCO$^+$                        & 9.3\,$\times$\,10$^{13}$ & \cite{Pratap1997} &  CH$_3$C$_3$N                   & 1.7\,$\times$\,10$^{12}$ & \cite{Marcelino2021} &  SO$_2$                         & 3.0\,$\times$\,10$^{12}$ & \cite{Cernicharo2011} \\
 H$_2$CO                        & 5.0\,$\times$\,10$^{14}$ & \cite{Ohishi1998} &  CH$_2$CCHCN                    & 2.7\,$\times$\,10$^{12}$ & \cite{Marcelino2021} &  NCS                            & 9.5\,$\times$\,10$^{11}$ & \cite{Cernicharo2024b} \\
 CH$_3$OH                       & 4.8\,$\times$\,10$^{13}$ & \cite{Cernicharo2020a} &  HCCCH$_2$CN                    & 2.8\,$\times$\,10$^{12}$ & \cite{Marcelino2021} &  HNCS                           & 3.2\,$\times$\,10$^{11}$ & \cite{Cernicharo2024b} \\
 C$_2$O                         & 7.5\,$\times$\,10$^{11}$ & \cite{Cernicharo2021f} &  $t$-CH$_3$CHCHCN               & 5.0\,$\times$\,10$^{10}$ & \cite{Cernicharo2022c} &  HSCN                           & 8.3\,$\times$\,10$^{11}$ & \cite{Cernicharo2024b} \\
 HCCO                           & 7.7\,$\times$\,10$^{11}$ & \cite{Cernicharo2021f} &  $c$-CH$_3$CHCHCN               & 1.3\,$\times$\,10$^{11}$ & \cite{Cernicharo2022c} &  HCNS                           & 9.0\,$\times$\,10$^{09}$ & \cite{Cernicharo2024b} \\
 CH$_2$CO                       & 1.4\,$\times$\,10$^{13}$ & \cite{Cernicharo2020a} &  CH$_2$C(CH$_3$)CN              & 1.0\,$\times$\,10$^{11}$ & \cite{Cernicharo2022c} &  CH$_3$CHS                      & 9.8\,$\times$\,10$^{10}$ & \cite{Agundez2025a} \\
 CH$_3$CO$^+$                   & 3.2\,$\times$\,10$^{11}$ & \cite{Cernicharo2021g} &  $g$-CH$_2$CHCH$_2$CN           & 8.0\,$\times$\,10$^{10}$ & \cite{Cernicharo2022c} &  C$_3$S                         & 6.8\,$\times$\,10$^{12}$ & \cite{Fuentetaja2025} \\
 CH$_3$CHO                      & 3.5\,$\times$\,10$^{12}$ & \cite{Cernicharo2020a} &  $c$-CH$_2$CHCH$_2$CN           & 7.0\,$\times$\,10$^{10}$ & \cite{Cernicharo2022c} &  HC$_3$S                        & 1.5\,$\times$\,10$^{11}$ & \cite{Cernicharo2024c} \\
 C$_2$H$_3$OH                   & 2.5\,$\times$\,10$^{12}$ & \cite{Agundez2021a} &  CH$_2$(CN)$_2$                 & 1.8\,$\times$\,10$^{11}$ & \cite{Agundez2024} &  HC$_3$S$^+$                    & 2.0\,$\times$\,10$^{11}$ & \cite{Cernicharo2021j} \\
 C$_2$H$_5$OH                   & 1.1\,$\times$\,10$^{12}$ & \cite{Agundez2023b} &  C$_5$N                         & 4.7\,$\times$\,10$^{11}$ & \cite{Agundez2023a} &  H$_2$C$_3$S                    & 7.3\,$\times$\,10$^{11}$ & \cite{Esplugues2026} \\
 CH$_3$OCH$_3$                  & 2.5\,$\times$\,10$^{12}$ & \cite{Agundez2021a} &  C$_5$N$^-$                     & 8.8\,$\times$\,10$^{10}$ & \cite{Agundez2023a} &  $c$-C$_3$H$_2$S                & 4.8\,$\times$\,10$^{10}$ & \cite{Esplugues2026} \\
 HCO$_2^+$                      & 4.0\,$\times$\,10$^{11}$ & \cite{Cernicharo2020a} &  HC$_5$N                        & 1.8\,$\times$\,10$^{14}$ & \cite{Cernicharo2020b} &  CH$_2$CHCHS                    & 4.4\,$\times$\,10$^{10}$ & \cite{Cabezas2025b} \\
 HCOOH                          & 9.3\,$\times$\,10$^{11}$ & \cite{Molpeceres2025} &  HC$_5$N$^+$                    & 9.9\,$\times$\,10$^{10}$ & \cite{Cernicharo2024d} &  HCSCN                          & 1.3\,$\times$\,10$^{12}$ & \cite{Cernicharo2021h} \\
 $c$-HCOOH                      & 5.3\,$\times$\,10$^{10}$ & \cite{Molpeceres2025} &  HC$_4$NC                       & 3.0\,$\times$\,10$^{11}$ & \cite{Cernicharo2020b} &  HCCCHS                         & 3.2\,$\times$\,10$^{11}$ & \cite{Cernicharo2021h} \\
\hline
\end{tabular}
\tablenotea{\\
Notes.-- Column densities can be converted to abundances relative to H$_2$ by dividing by 10$^{22}$ cm$^{-2}$ (Ref.~\citenum{Cernicharo1987}). $^a$ Tentative detection.
}
\end{table}

To address the above question we have taken the chemical network of the latest release of the UMIST Database for Astrochemistry \cite{Millar2024} and run a standard pseudo time-dependent gas-phase chemical model of a cold dark cloud \cite{Agundez2013}. We adopt a temperature of 10 K, a volume density of \ch{H2} molecules of 10$^4$ cm$^{-3}$, a cosmic-ray ionization rate of \ch{H2} of 1.3\,
$\times$\,10$^{-17}$ s$^{-1}$, a visual extinction of 30 mag, and the so-called set of 'low-metal' elemental abundances \cite{Agundez2013}. A way to evaluate the goodness of the chemical model is to quantify the global level of agreement between calculated and observed abundances. To this end we use the distance of disagreement \cite{Wakelam2006}, which is defined as
\begin{equation}
D = \frac{\sum_{i=1}^n |\log_{10} (f_i^{cal}) - \log_{10} (f_i^{obs})|}{n},
\end{equation}
where $f_i^{cal}$ and $f_i^{obs}$ are the calculated and observed fractional abundances of molecule $i$ and the sum runs over all $n$ observed molecules that are included in the chemical model, which in our case is 142. The 46 molecules missing in the UMIST network are mainly isomers of already included species, such as cis HCOOH \citep{Molpeceres2025}, and PAHs like phenalene \citep{Cabezas2025c}. Indeed, isomers and PAHs are probably the two main types of molecules missing in state-of-the-art chemical models. To evaluate how sensitive is the agreement between calculated and observed abundances to the elemental C/O ratio, we let it to increase from the solar value to values above one by decreasing the abundance of oxygen while the abundance of carbon is kept fixed to 1.8\,$\times$\,10$^{-4}$ relative to H. Variations in the C/O ratio were proposed to explain the observed abundance gradients in the dark clouds L134N \cite{Swade1989} and TMC-1 \cite{Pratap1997}. Indeed, an enhancement in the C/O over the solar value is known to boost the abundances of C-bearing molecules such as carbon chains and PAHs, resulting in a better agreement with the composition of TMC-1 \cite{Terzieva1998,Wakelam2006,Agundez2013,Byrne2026}. The scientific rationale behind decreasing the abundance of oxygen is based on the observational finding of depletion of oxygen on dust with increasing cloud density \cite{Jenkins2009,Hincelin2011}.

\begin{figure}
\centering
\includegraphics[angle=0,width=0.88\textwidth]{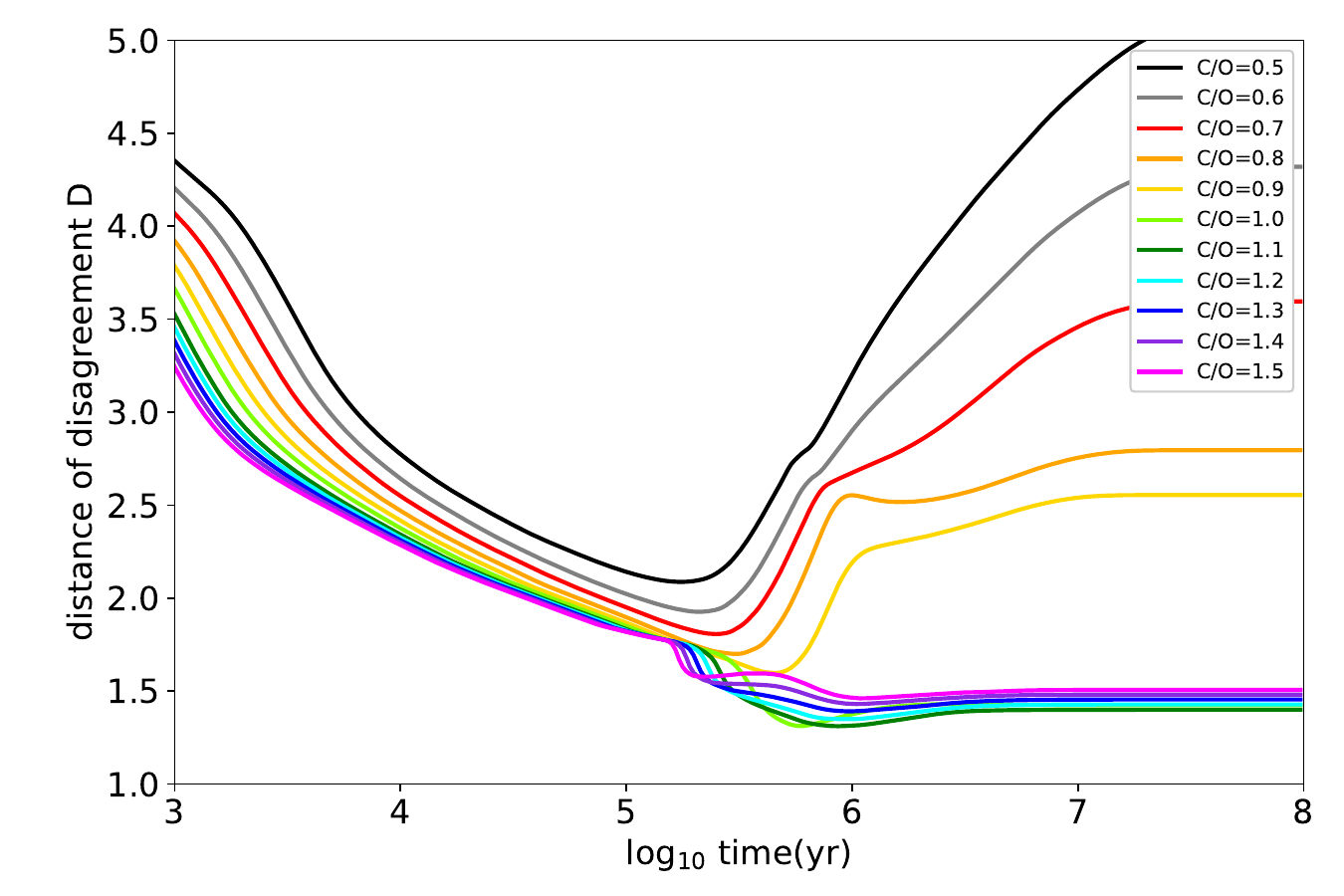}
\caption{{\small Distance of disagreement D calculated as a function of time for different elemental C/O ratios. The lower is D the better is the agreement between the chemical model and the observations.}} \label{fig:d}
\end{figure}

In Fig.\,\ref{fig:d} we show the distance of disagreement as a function of time for various C/O ratios. The best agreement is reached at times between 10$^5$ and 10$^6$ yr, although the exact time and the level of agreement reached depends heavily on the C/O ratio. This range of times is consistent with the expected lifetime of a starless core \cite{Offner2025}. The agreement between the chemical model and observations improves smoothly as the C/O increases from a solar value, around 0.5, and saturates once the C/O ratio reaches unity. The situation could be different if one looks at individual molecules instead of at the global composition. For example, in the case of PAHs such as naphthalene increasing the C/O ratio above one improves the agreement with observations \cite{Byrne2026}, although too high C/O ratios would cause a severe overestimation of carbon chains \cite{Agundez2013}. It is clear that C/O ratios around or above one are preferred. Whether such a C/O ratio is realistic for TMC-1 and what would be the ultimate cause of the carbon enrichment of the gas phase, either oxygen depletion or another, is yet an open question. It is interesting to note that the distance of disagreement never falls below one. Concretely, the smallest D is 1.3 and it is reached for the model with C/O = 1 at a time of $\sim$\,5\,$\times$\,10$^5$ yr. That is, the average level of disagreement between calculated and observed abundances if larger than one order of magnitude, which illustrates well the margin for improvement of state-of-the-art chemical networks in use in interstellar chemistry.

\begin{figure}[t]
\centering
\includegraphics[angle=0,width=0.88\textwidth]{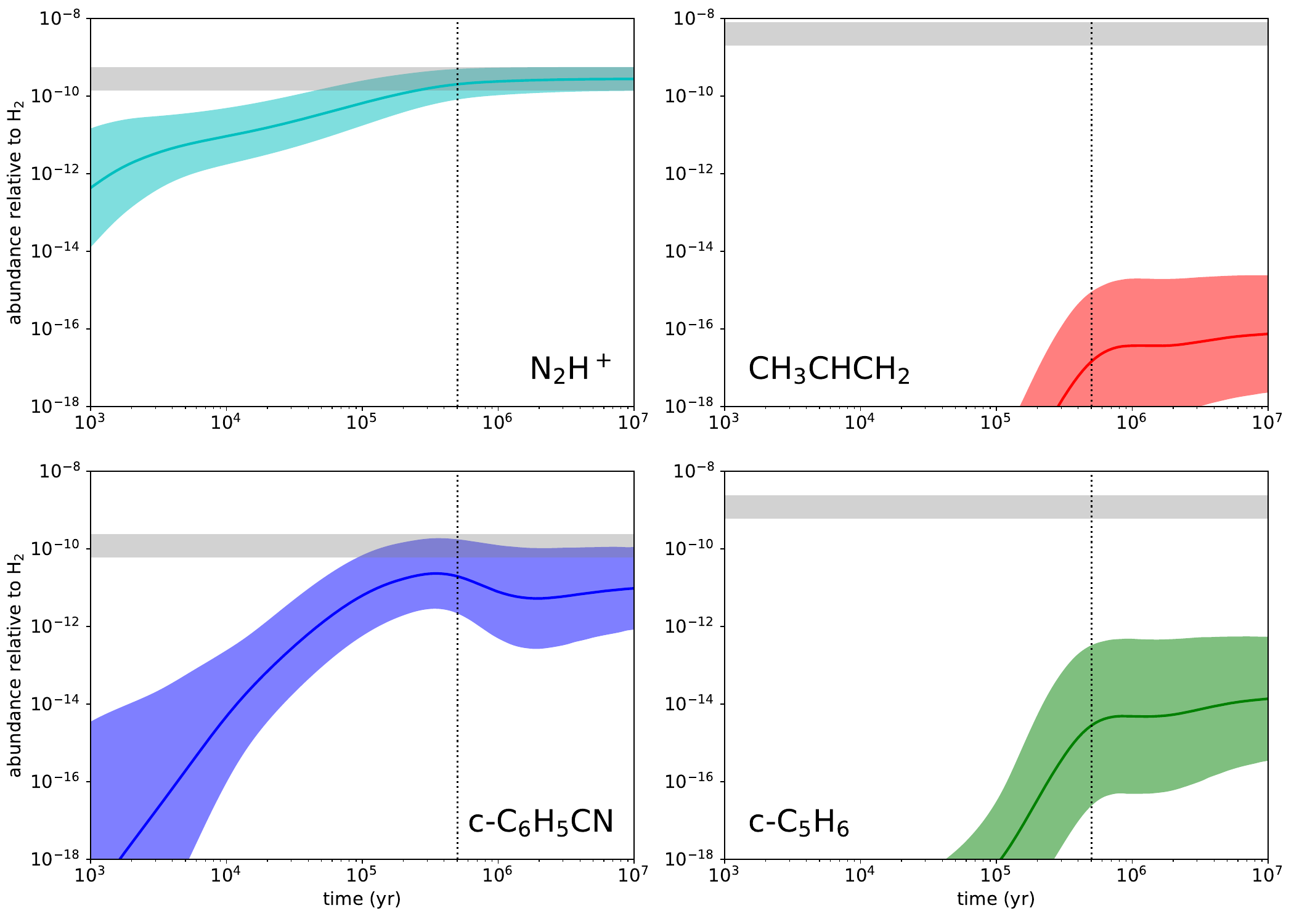}
\caption{{\small Calculated and observed fractional abundances of N$_2$H$^+$, CH$_3$CHCH$_2$, c-C$_6$H$_5$CN, and c-C$_5$H$_6$. The colored areas indicate the mean calculated abundance with the corresponding uncertainty, while the gray horizontal band indicate the observed abundance with the corresponding error, estimated to be a factor of two. The vertical dotted line corresponds to the time of best global agreement between calculated and observed abundances, 5\,$\times$\,10$^5$ yr.}} \label{fig:abundances}
\end{figure}

Although the global goodness of the UMIST network for TMC-1 is at the level of one order of magnitude, the level of agreement is very different depending on the molecule. In Fig.\,\ref{fig:abundances} we compare calculated and observed abundances for four selected molecules. The calculated abundances were computed by running 1000 chemical models in which the reaction rate coefficients were randomly varied within their uncertainties using a lognormal distribution \cite{Wakelam2005}. We took advantage of the fact that the UMIST network contains estimated uncertainties for reaction rate coefficients \cite{Millar2024}. In this series of runs we adopted a elemental C/O ratio of 1. A similar sensitivity analysis focused on the identification of critical reactions in TMC-1 has been recently carried out \cite{Byrne2024}, in that case using the Nautilus code \cite{Ruaud2016} which in turn uses the KIDA network \cite{Wakelam2024}. In the case of the cation N$_2$H$^+$, the agreement between calculated and observed abundance is very good. In addition, the uncertainty in the calculated abundance is below a factor of three for times longer than 10$^5$ yr. From these two facts we can conclude that the chemistry of N$_2$H$^+$ in cold dark clouds is well understood. If we switch to the case of propene (\ch{CH3CHCH2}), the calculated abundance falls various orders of magnitude below the observed value, which indicates that the chemical network is missing important formation routes to this hydrocarbon. In addition, the uncertainty in the calculated abundance is large, between one and two orders of magnitude. The formation of propene in dark clouds continues to be a remarkable open problem in astrochemistry \cite{Lin2013,Millar2024}. The hydrocarbon cycles benzonitrile and cyclopentadiene are in two very distinct situations. In the case of benzonitrile, there is an acceptable agreement between calculated and observed abundance and the uncertainty in the calculated abundance lies at the level of one order of magnitude. However, for cyclopentadiene the chemical model severely underestimates the abundance, which in turn has a sizable uncertainty of around two orders of magnitude. It is clear that we still do not understand well how is cyclopentadiene formed in cold dark clouds. In the case of benzonitrile, and its most likely precursor benzene, there appears to be efficient formation routes, although some of them are still in question, as will be discussed in next section.

\section{The origin of PAHs} \label{sec:pahs}

The chemical content of TMC-1 and similar cold dark clouds has been long thought to be characterized by the presence of long carbon chains of the family of polyynes and cyanopolyynes. This characteristic has been largely probed by radioastronomical observations \cite{Kroto1978,Suzuki1986,Kalenskii2004,Loomis2021,Remijan2023} and explained in terms of gas-phase synthesis by chemical models, most of which have been largely carried out by Prof. Eric Herbst or inspired by his work \cite{Leung1984,Herbst1986a,Herbst1986b,Millar1988,Herbst1989,Herbst1994,Winstanley1996,Smith2004,Wakelam2006,Wakelam2008,Walsh2009}. In this context, one of the most surprising discoveries made in recent years has been the detection in TMC-1 of a very abundant population of large cyclic structures of the family of polycyclic aromatic hydrocarbons. The story started with the detection of benzonitrile, first tentatively \cite{Kalenskii2017} and then confirmed \cite{McGuire2018}. Benzonitrile (\ch{C6H5CN}) acts as a proxy of the radioinvisible molecule benzene \cite{Cooke2020}, which is an iconic molecule in organic chemistry and the doorway to larger aromatic molecules. Later on, other aromatic cycles were also detected in TMC-1\footnote{Not all the cyclic molecules detected in TMC-1 are strictly aromatic. For example, cyclopentadiene is clearly not aromatic and molecules composed of fused six- and five-membered rings such as indene are aromatic in the part of the six-membered rings but not in the five-membered one. We nevertheless embrace cyclopentadiene and larger cyclic molecules into the family of aromatic cycles because all them are probably chemically linked.}. These comprise molecules consisting of one ring such as cyclopentadiene (\ch{C5H6}) \cite{Cernicharo2021d}, two fused rings such as indene (\ch{C9H8}) \cite{Cernicharo2021d,Burkhardt2021a} and cyanonaphthalene (\ch{C10H7CN}) \cite{McGuire2021}, three fused rings such as cyanoacenaphthylene (\ch{C12H7CN}) \cite{Cernicharo2024e,Cernicharo2026a}, phenalene (\ch{C13H10}) \citep{Cabezas2025c}, and cyclopentindene (\ch{C11H8}) \citep{Fuentetaja2026}, four rings such as cyanopyrene (\ch{C16H9CN}) \cite{Wenzel2024,Wenzel2025a}, and even a molecule with seven fused rings such as cyanocoronene (\ch{C24H11CN}) \cite{Wenzel2025b}. Some of these aromatic cycles have also been found in other cold dark clouds similar to TMC-1 \cite{Burkhardt2021b,Agundez2023d}, which indicates that PAHs are common in clouds at this particular evolutionary stage of cold starless cloud.

The discovery of the aforementioned cycles constitute the first unequivocal detection of PAHs in the interstellar medium, although the discussion on the presence of PAHs in space has a long tradition in astronomy. Infrared (IR) emission bands observed in a wide variety of astronomical environments are commonly thought to arise from large PAHs with 20-100 carbon atoms \cite{Leger1984,Allamandola1985,Tielens2008}. However, no individual molecule of the family of PAHs had been unambiguously detected in the interstellar medium prior to the observations of TMC-1. In fact, there is still debate on the exact nature of the carriers of these IR bands, which may have a mixed aromatic-aliphatic character \cite{Kwok2011}. IR emission bands are observed in almost every astronomical region where there is an appreciable UV radiation field, such as planetary nebulae, dense PDRs in star-forming regions, protoplanetary disks, and diffuse clouds. UV light is required to excite PAHs and produce emission at IR wavelengths through fluorescence \cite{Tielens2008}. Cold dark clouds, which are protected against external UV light, were therefore not envisaged to bring the first indisputable evidence of the presence of PAHs in space, although they are favored for radioastronomical detection due to the low temperatures and thus low partition functions. The regions showing IR emission bands are connected evolutionary in the sense: evolved stars $\rightarrow$ diffuse clouds $\rightarrow$ dense clouds $\rightarrow$ protostellar cores $\rightarrow$ protoplanetary disks $\rightarrow$ planetary systems. Therefore, the population of large PAHs is likely to arise in the ejecta of evolved stars (still unclear whether in the AGB or planetary nebula stages) and would survive along the way to planetary systems, although probably being subject to some degree of processing at each particular evolutionary stage. It seems logical to think that large PAHs are present inside cold dark clouds inherited from the previous evolutionary phase of diffuse cloud \cite{Wakelam2008}. The lack of IR emission bands from dark clouds would merely be a consequence of the unavailability of UV radiation to excite large PAHs. So the question is: is the population of relatively small (5-24 carbon atoms) PAHs detected through radioastronomical observations in cold dark clouds such as TMC-1 connected to the population of large (20-100 carbon atoms) PAHs probed through IR observations that pervades all evolutionary stages? A possible link would be that small PAHs originate from the fragmentation of preexisting large PAHs in a top-down scenario. An alternative hypothesis is that small PAHs are formed from small hydrocarbons in a bottom-up scheme. We discuss below the possible origin of small PAHs in cold dark clouds, which is still a major open question in astronomy.

There are several observational pieces of evidence related to benzonitrile that favor a bottom-up scenario. First, the spatial distribution of the aromatic molecule benzonitrile in TMC-1 resembles that of cyanopolyynes \cite{Cernicharo2023b} (see Fig.\,\ref{fig:maps}). Second, observations of several dark clouds indicate that the column density of this same molecule, \ch{C6H5CN}, is positively correlated with that of the cyanopolyyne \ch{HC7N}, which has the same number of heavy atoms \cite{Agundez2023d}. That is, the higher the abundance of \ch{HC7N}, the more abundant \ch{C6H5CN} is. These two observational findings suggest that the chemical processes that are responsible for the formation of carbon chains, which are formed in situ through bottom-up routes, are also behind the synthesis of aromatic cycles in dark clouds. Third, the low D/H ratio inferred for \ch{C6H5CN} in TMC-1 \cite{Steber2025} is in line with the range of D/H ratios observed for other molecules, which are explained in terms of in situ isotopic fractionation at low temperature, but it is lower than the range of D/H ratios observed for large PAHs in PDRs, where PAHs are known to be enriched in D due to mechanisms not yet fully understood \cite{Allamandola1989,Peeters2024}. This result suggests that small PAHs in dark clouds are unlikely to result from the fragmentation of larger PAHs in a top-down process. Even if there are evidences that favor a bottom-up scenario, it is by no means clear which are the chemical reactions responsible of the synthesis. There are several possibilities, either based on neutral-neutral reactions or involving ion-neutral reactions, that are discussed below. 

\begin{figure}
\centering
\includegraphics[angle=0,width=0.90\textwidth]{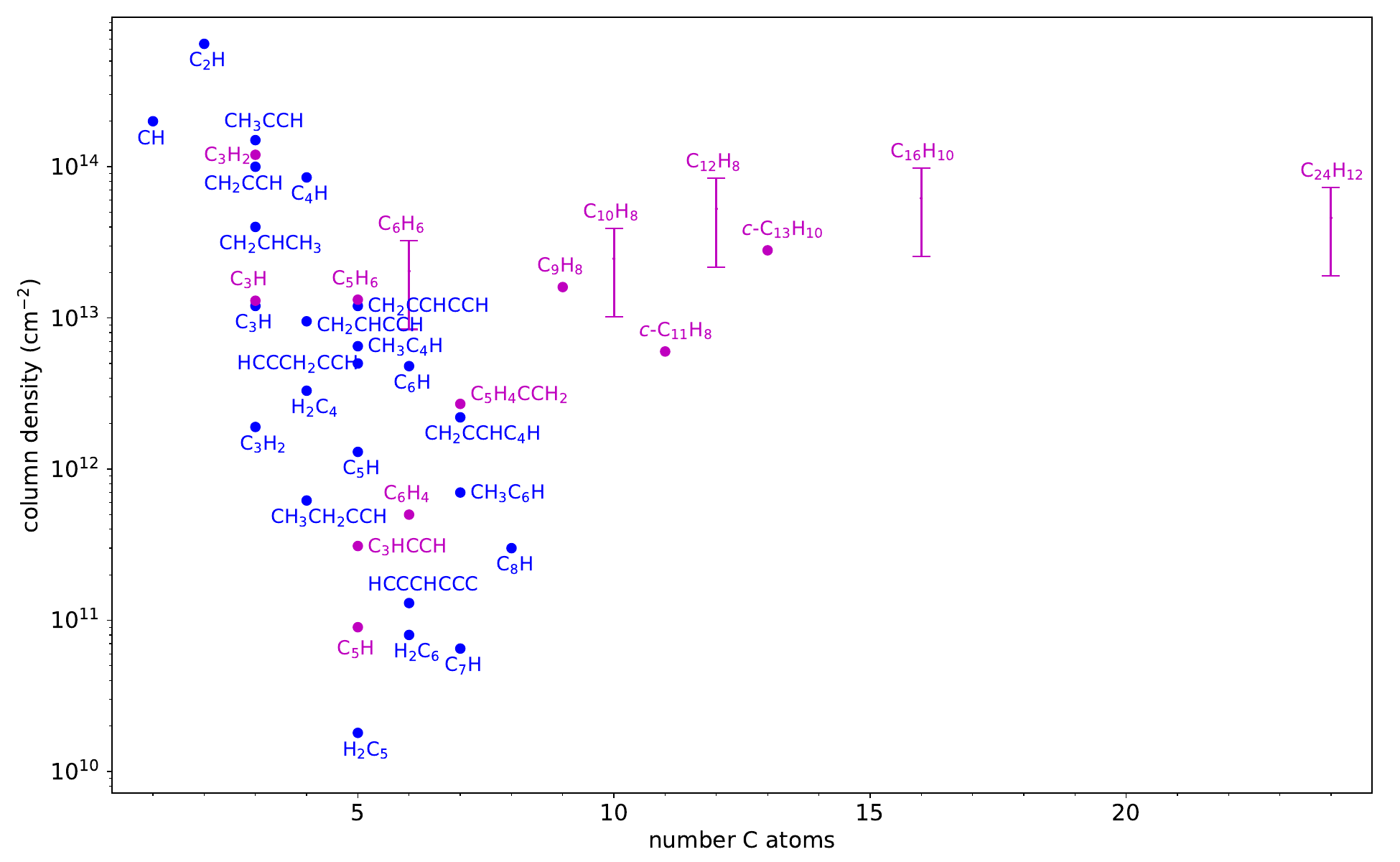}
\caption{{\small Column densities of neutral hydrocarbons in TMC-1 as a function of the number of carbon atoms. Cyclic species are highlighted in color magenta. The ranges of column densities of non-polar or nearly non-polar cycles are estimated from those of the cyano derivatives (see text).}} \label{fig:hydrocarbon}
\end{figure}

Whatever it is the synthetic pathway to aromatic cycles in dark clouds, the reactions involved on it must be chemically feasible, i.e., they must be fast at low temperature and produce with a significant yield the right products, as well as astronomically feasible, that is, the reactants involved must be abundant enough in dark clouds. In the end, whether a given chemical scheme of synthesis is functional at the conditions of dark clouds must be validated by a chemical model. However, we are still far from having at hand a full chemical mechanism able to describe the synthesis of PAHs in cold clouds, and only the chemical and astronomical feasibility of particular individual reactions has been investigated. One important point to have in mind is that observed aromatic cycles in TMC-1 have very large abundances. In Fig.\,\ref{fig:hydrocarbon} we represent the column densities of neutral hydrocarbons in TMC-1 as a function of the molecular size. We note that the column densities of the non-polar or nearly non-polar cycles \ch{C6H6}, \ch{C10H8}, \ch{C12H8}, \ch{C16H10}, and \ch{C24H12} are estimated from those of the observed cyano derivatives assuming an abundance ratio between the pure hydrocarbon and the $-$CN derivative in the range 7-27, where the lower bound is an estimation for coronene based on chemical kinetics arguments on the reaction between CN and the corresponding unsubstituted hydrocarbon \cite{Wenzel2025a,Wenzel2025b} and the upper bound corresponds to the ratio observed for cyclopentadiene in TMC-1 \cite{Cernicharo2021d,Cernicharo2021e}. We cannot neglect the possibility that cyano derivatives are also formed by alternative routes to that of CN reacting with the unsubstituted hydrocarbon, although the current body of evidence points to CN + hydrocarbon being the major pathway. In the case of benzene, its reaction with CN is known to be fast at low temperatures \citep{Cooke2020}. The reaction between cyano 1,3-butadiene and C$_2$H could also form benzonitrile (Fig.\,D.1 in \citealt{Cernicharo2021e}) but the low abundance of cyano 1,3-butadiene in TMC-1 \citep{Cooke2023,Agundez2025b} makes it unlikely that this route is an important one. In any case, a better knowledge of the role of alternative routes to CN derivatives would require to measure the kinetics of reactions between CN and different PAHs and to expand the number of cases in which the unsubstituted and CN-substituted are both detected. It is striking that all species of the family of PAHs lie at the same level of abundance, with column densities in the range 10$^{13}$-10$^{14}$ cm$^{-2}$, and that this level is nearly independent of the size and quite high, comparable or slightly lower than that of the most abundant hydrocarbons with 1-4 carbon atoms, such as \ch{CH}, \ch{C2H}, c-\ch{C3H2}, \ch{CH2CCH}, \ch{CH3CCH}, and \ch{C4H}, all of which are potential precursors of PAHs. We note that although there are non-negligible uncertainties in the abundances of PAHs, in particular for those derived from CN derivatives, it is unlikely that this could modify the observed pattern of nearly uniform abundance of PAHs independently of their size. This fact imposes a strong constraint on any potential scheme of synthesis, which must be extraordinarily efficient to convert small hydrocarbons into very large ones without a drastic loss of abundance. This is in contrast to the case of carbon chains, where the growth occurs at the expense of a significant drop in abundance as the chain length increases (see, e.g., the sequence \ch{C2H} $\rightarrow$ \ch{C4H} $\rightarrow$ \ch{C6H} $\rightarrow$ \ch{C8H} in Fig.\,\ref{fig:hydrocarbon}).

There are several neutral-neutral reactions that could play a role in the synthesis of aromatic rings. The self-reaction of propargyl, which is one of the most abundant radicals in TMC-1 \cite{Agundez2021b,Agundez2022a}, probably occurs with cyclization at low temperature, yielding \ch{C6H5} + H \cite{Zhao2021}. However, phenyl is a rather unreactive radical and it is not straightforward how it would lead to benzene. Our colleague, R. I. Kaiser, has proposed a chemical scheme based on chemically feasible neutral-neutral reactions (see Appendix D in \citealt{Cernicharo2021e}), where 1,3-butadiene would react with CH to yield cyclopentadiene \cite{He2020}, which upon further reaction with the radical CH would produce benzene \cite{Caster2021}. In this scheme, benzene could also be formed in the reaction of 1,3-butadiene with the radical C$_2$H \citep{Jones2011}. While this scheme appears to be chemically plausible, it is most likely not astronomically feasible because the key precursor of the whole scheme, 1,3-butadiene, is not abundant enough in TMC-1, as indicated by the non-detection of the 1-cyano substituted derivatives \cite{Agundez2025b}. Other neutral-neutral reactions found to be chemically feasible are the reaction between the propargyl (\ch{CH2CCH}) and vinyl (\ch{C2H3}) radicals, which would form the cyclopentadienyl radical \cite{GarciadelaConcepcion2023}, or the reaction of vinyl benzene and the radical CH, which would form indene \cite{Doddipatla2021}. It is yet to be evaluated whether these reactions are astronomical relevant in the conditions of TMC-1. Therefore, it is difficult to envisage neutral-neutral reactions that could efficiently form cycles such as cyclopentadiene and benzene in dark clouds. For example, the reaction between propene (\ch{CH2CHCH3}) and the radical \ch{C2H}, two fairly abundant hydrocarbon species in TMC-1 \cite{Marcelino2007,Sakai2010}, has been measured to be fast at low temperatures \cite{Chastaing1998}, although it forms acyclic \ch{C5H6} isomers rather than cyclopentadiene \cite{Goettl2022}. 

Ion-neutral reactions offer a more efficient means to form aromatic rings than neutral-neutral ones because they are usually much faster. For example, benzene is thought to form efficiently at dark cloud conditions through several ion-neutral reactions that lead to the C$_6$H$_5^+$ ion, which in turn associates radiatively with \ch{H2} to form C$_6$H$_7^+$, ultimately leading to benzene after dissociative recombination with electrons \citep{McEwan1999,Woods2011,Agundez2021b,Byrne2024}. In our chemical model, which uses the UMIST\,2022 network, the formation of benzene occurs through this sequence of reactions
\begin{equation}
\rm CH_4 + C_5H_2^+ \rightarrow \textnormal{c-}C_6H_5^+ + H,
\end{equation}
\begin{equation}
\rm \textnormal{c-}C_6H_5^+ + H_2 \rightarrow \textnormal{c-}C_6H_7^+ + \textit{h}\nu,
\end{equation}
\begin{equation}
\rm \textnormal{c-}C_6H_7^+ + e^- \rightarrow \textnormal{c-}C_6H_6 + H.
\end{equation}
Once benzene is formed, its reaction with CN would directly lead to benzonitrile \cite{Cooke2020}. Although the abundance calculated for benzonitrile agrees relatively well with the value inferred from observations (see Fig.\,\ref{fig:abundances}), there are important uncertainties in the above scheme. The radiative association between c-\ch{C6H5+} and \ch{H2} to yield c-\ch{C6H7+} has been put in question recently \cite{Kocheril2025} and there is debate on whether or not this reaction occurs \cite{Loison2025}. In addition, the branching ratios of the dissociative recombination of $c$-C$_6$H$_7^+$ with electrons, which is thought to lead to benzene, are not known. The situation is even worse for cyclopentadiene, in which case the calculated abundance lies various orders of magnitude below the observed value (see Fig.\,\ref{fig:abundances}). It has been proposed that l-\ch{C3H3+} and \ch{C2H4} could associate radiatively to yield c-\ch{C5H7+}, which upon dissociative recombination with electrons could finally lead to cyclopentadiene \cite{Cernicharo2022b}. There is experimental evidence that the reaction l-\ch{C3H3+} + \ch{C2H4} is fast and the ion \ch{C5H7+} has been identified as product, although it is not known whether or not the product \ch{C5H7+} is cyclic \cite{Smyth1982,Anicich2003}. However, recent theoretical calculations \cite{Mallo2025} find that the preferred products are cyclic C$_5$H$_5^+$ + H$_2$ rather than C$_5$H$_7^+$, which would make this reaction not viable to synthesize cyclopentadiene.

If we accept that large PAHs inherited from the previous stage of diffuse cloud are present in cold dark clouds, this would provide an abundant enough reservoir to act as precursor of smaller cycles. It is however not straightforward how these large cycles would fragment to yield smaller rings. The UV flux inside dark clouds is small \cite{Prasad1983}. Cosmic rays could drive this fragmentation, although the typical ionization rates caused by cosmic rays, on the order of 10$^{-17}$ s$^{-1}$, result in low yields and the fragments tend to be chains rather than cycles \cite{Chabot2020}. Collisional fragmentation is also unlikely given the low temperature and quiescent nature of the cloud, as evidenced by the small line widths.

In addition to the problem of PAH formation, there are also problems associated to the presence itself of large quantities of PAHs in the gas phase of cold clouds such as TMC-1. The question is how is it possible that such large molecules survive in the gas phase and have not been deposited on dust grains as ices, taking into account the low temperature of the cloud and the large cross section of the molecules. The distribution of PAHs in the gas and ice is regulated by the balance between adsorption and desorption, where sputtering induced by cosmic rays would be the main desorption mechanism in UV-shielded cold clouds \cite{Dartois2022}. In these conditions, for PAHs such as as indene and naphthalene the number of molecules on ices is expected to greatly exceed that in the gas phase. Taking into account the large abundance of PAHs such as indene and naphthalene observed in the gas phase of TMC-1, the estimated amount in the form of ices is unrealistically large \cite{Mate2023}.

\section{Concluding remarks}

Thanks to the observational efforts dedicated to characterize the chemical composition of TMC-1 by the QUIJOTE and GOTHAM projects, to date the inventory of molecules known to be present in this cold dark cloud amounts to 188. Among them there is a vast variety of organic molecules comprising carbon chains and aromatic cycles. State-of-the-art chemical models developed upon the pioneering work of Prof. Eric Herbst describe reasonable well the observed abundances of many molecules but cannot yet account for the formation of aromatic molecules such as PAHs. Further work is needed to quantitatively assess the feasibility of bottom-up and top-down processes to explain the origin of this remarkable family of molecules.

\begin{acknowledgement}

We would like to express our gratitude to Prof. Eric Herbst for being an immense source of inspiration during all our scientific career and for his enthusiasm, open mind, and friendly character. We also thank the three anonymous referees for insightful comments that helped to improve this manuscript. We acknowledge funding support from Spanish Ministerio de Ciencia, Innovaci\'on, y Universidades through grant PID2023-147545NB-I00 and from European Research Council through grant ERC-2013-Syg-610256-NANOCOSMOS. We thank the computational resources provided by the DRAGO computer cluster managed by SGAI-CSIC, and the Galician Supercomputing Center (CESGA).

The authors declare no competing financial interests.

\end{acknowledgement}

\bibliography{references}

\end{document}